\documentclass[preprint,amsmath,amssymb,superscriptaddress]{revtex4}
\usepackage{graphicx,epstopdf}

\newcommand{\Fkt}[1]{\,\mathsf {#1}}
\ifx\Tr\renewcommand{\Tr}{\Fkt{Tr}}
\else\newcommand{\Tr}{\Fkt{Tr}}
\fi

\begin{document}

\title{Photoassociation spectroscopy in Penning ionization reactions
  at sub-Kelvin temperatures}

\author{Wojciech Skomorowski}
\affiliation{Theoretische Physik, Universit\"at Kassel,
  Heinrich-Plett-Stra{\ss}e 40, 34132 Kassel, Germany
}

\author{Yuval Shagam}
\affiliation{Department of Chemical Physics, Weizmann Institute of Science,
  Rehovot 76100,  Israel}

\author{Edvardas Narevicius}
\affiliation{Department of Chemical Physics, Weizmann Institute of Science,
  Rehovot 76100,  Israel}

\author{Christiane P. Koch}
\affiliation{Theoretische Physik, Universit\"at Kassel,
  Heinrich-Plett-Stra{\ss}e 40, 34132 Kassel, Germany
}
\email{christiane.koch@uni-kassel.de}

\begin{abstract}
Penning ionization reactions in merged beams with precisely controlled
collision energies have been shown to accurately probe quantum mechanical
effects in reactive collisions. A complete microscopic understanding
of the reaction is, however, faced with two major challenges---the
highly excited character of the reaction's entrance channel and the
limited precision of even the best state-of-the-art \textit{ab initio}
potential energy surfaces. Here, we suggest photoassociation
spectroscopy as a tool to identify the character of orbiting
resonances in the entrance channel and probe the ionization width as a
function of inter-particle separation. We introduce the basic concept
and discuss the general conditions under which this type of
spectroscopy will be successful. 
\end{abstract}
\date{\today}

\maketitle

\section{Introduction}
Understanding chemical reaction dynamics in  a microscopic
perspective from first principles has been a long standing goal in
physical chemistry. Over many decades, molecular beams have been the
prime tool to study gas phase reactions. Controlling the beams with
electric or magnetic fields has recently rejuvenated this 
field~\cite{vandeMeerakkerCR12,vanBuurenPRL09,NareviciusCR12}. 
In particular, merging two beams with a magnetic
field~\cite{HensonSci12,Lavert-OfirNatChem14,BertscheChimia14,JankunasJCP14,JankunasJCP15,ShagamNatChem15}
has allowed for 
unprecedented control over the collision energy, both in terms of
accessible range and precision.
Studying Penning ionization reactions with this technique comes with
the additional advantage of nearly unit efficiency in detecting the
ionic reaction products.
This has allowed for observing orbiting resonances
at sub-Kelvin temperatures~\cite{HensonSci12}
including an isotope effect~\cite{Lavert-OfirNatChem14}
and elucidating the role of internal rotation for the
reaction~\cite{ShagamNatChem15}. 
However, a complete microscopic understanding of the reaction dynamics
requires high-level \textit{ab initio} theory in addition to precise
experimental data. While the entrance
channel of the reaction can be described with reasonable
accuracy~\cite{HapkaJCP13},  
a full model of Penning ionization reactions is faced with two major
challenges. First, since it necessarily involves metastable states, 
the entrance channel is a highly excited electronic state, which 
is embedded in a continuum of ionized states.
The corresponding electronic wave function does not belong to the space
of square integrable functions  and thus  cannot be correctly    
described by standard electronic structure  methods, designed for the 
description of bound states.
The  difficulty of coupling to a continuum of ionized states can be
circumvented by replacement with an imaginary potential, 
defined within the Fano-Feshbach formalism
for an isolated resonance~\cite{Fano, Feshbach}. In this approach, a
projection operator is introduced, which allows to partition the
total Hilbert space into discrete 
and continuum parts. The imaginary potential is obtained 
from the coupling matrix elements between wave  functions in 
the two subspaces. However, choice and construction of the 
Feshbach projection operator are highly nontrivial. 
Second, even for diatomics, the best, currently available potential
energy curves are 
not sufficiently accurate to correctly predict the scattering length
and thus  exact position and character 
of orbiting resonances~\cite{LondonoPRA10}. 

Here, we suggest photoassociation spectroscopy of Penning ionization
reactions to address these two issues. Photoassociation refers to the
formation of a chemical bond upon laser
excitation~\cite{FrancoiseReview,JonesRMP06}. 
In standard photoassociation spectroscopy, the laser excites two
colliding ground state atoms to an electronically excited
state. Detuning the laser frequency from the atomic excitation energy
probes molecular levels below the excited state dissociation
threshold~\cite{FrancoiseReview,JonesRMP06}. 
Photoassociation rates are limited by the amplitude of
collision wavefunctions at the Condon radius, i.e., the inter-particle
separation where the laser frequency matches the energy difference
between ground and excited potential energy surface~\cite{KochJPhysB06}. 
At sub-Kelvin temperatures, only few partial waves
contribute to the collision and, due to the rotational barrier and
quantum reflection, their
wavefunctions have little amplitude at short inter-particle
separations. Photoassociation is thus most effecient at intermediate
and large separations~\cite{KochCR12}.
This implies excitation into weakly bound levels just below the
dissociation threshold.  

\begin{figure}[tb]
   \includegraphics[width=0.5\linewidth, angle=-90]{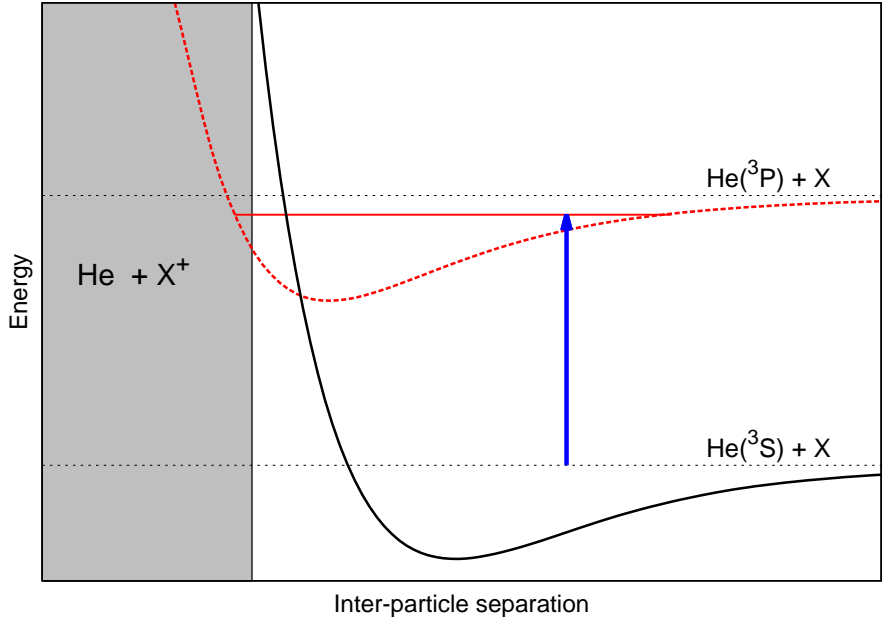}
   \caption{Photoassociation spectroscopy in Penning ionization. The
     laser, indicated by the blue arrow, excites a scattering pair
     colliding with collision energy $E$ in the He$(^3S)$+X state 
     into a bound rovibrational level in the potential energy curve
     dissociating into He$(^3P)$+X. The grey shaded area at short
     inter-particle separations indicates the region where the
     probability for Penning ionization is unity. If the bound level
     extends into this region, ionization will happen faster than in
     the entrance channel, i.e., photoassociation will be obserevd in
     terms of an increased ionization rate. 
  }
  \label{fig0}
\end{figure}
When applied to a collision complex undergoing a Penning ionization
reaction, the electronic ground state as  entrance channel is replaced
by a metastable one, see Fig.~\ref{fig0}. Such a configuration of
electronic states can also be used to produce metastable diatomic
helium molecules via photoassociation~\cite{LeonardPRL03}. While in 
that case, the giant bond length of the helium dimer precludes Penning
ionization, here we consider interactions where the particles approach
each other close enough to Penning ionize. 
As specific example, we consider
photoassociation in the Penning ionization of metastable helium and
argon. However, our considerations hold also when argon is replaced
by a different ground state scattering partner $X$.
Penning ionization in collisions between He($^3S$)  and Ar was
extensively studies in the past,  see Ref.~\cite{SiskaRMP93} and
references therein; the most recent state-of-the-art experimental  
investigation used  velocity controlled merged
beams~\cite{HensonSci12}. As indicated by the blue arrow in
Fig.~\ref{fig0}, the  laser is taken to be red-detuned with respect to
the $^3P \; \leftarrow {^3S}$ line of helium, which is a
dipole-allowed  transition,  
\[
\mathrm{He}(^3S)+\mathrm{Ar}\xrightarrow{\;\; h\nu\;\;} 
\mathrm{He}(^3P)\mathrm{Ar}\,.
\]
Excitation then happens into bound levels below the He$(^3P)$+Ar
threshold. Since the excited state potential energy curve extends to
shorter inter-particle separation, where the probability for Penning
ionization is significantly increased, the reaction rate is larger
than in the entrance channel. Such a configuration of potential energy
curves has recently been reported for Penning ionization in
He($^3P$)+H$_2$ compared to that in He($^3S$)+H$_2$, resulting in 
orders of magnitude larger reaction
rates~\cite{ShagamNatChem15}. Photoassociation should 
thus be measurable by an enhancement of the Penning ionization
rate. However, observation of the bound levels is possible only, 
if the Penning ionization reaction is not too fast. More
quantitatively, the inverse lifetime of the 
bound levels due to Penning ionization needs to be smaller than the
level spacing. We will discuss the prospects for photoassociation
spectroscopy for the example of the Penning ionization of metastable 
helium and argon, based on a first principles model.

The remainder of the paper is organized as follows: We introduce our
model and the computational details in Section~\ref{sec:model} and
present our results in Section~\ref{sec:results}. In particular, we
discuss prospects to use photoassociation spectroscopy as a means to
unequivocally identify the rotational quantum number of orbiting
resonances and as a tool to map out the imaginary potential governing
the ionization probability. We conclude in Section~\ref{sec:concl}. 

\section{Model and computational details}
\label{sec:model}
The interaction between two particles undergoing a Penning ionization
reaction can be modeled be means of a complex potential~\cite{SiskaRMP93}, 
\begin{equation}
  W(R) = V(R) +\frac{i}{2} \,  \Gamma(R)   \,,
\end{equation}
where $R$ denotes inter-particle separation. 
$V(R)$ is the real part of the potential  and  $\Gamma(R)$ 
its width, also called imaginary or optical potential. In a
semi-classical picture, $V(R)$  governs the collisional dynamics of
the interacting species  while $\Gamma(R)$ reflects the probability
for Penning ionization as 
a function of inter-particle  distance. Both $V(R)$  and
$\Gamma(R)$ depend  on the electronic state, on which the interaction takes
place. 

In our model of Penning ionization of metastable helium and argon, 
three non-relativistic electronic states are relevant:
$(1)^3\Sigma^+$, asymptotically dissociating into He($^3S$) + Ar
atoms, and $(2)^3\Sigma^+$ 
and $^3\Pi$, dissociating into He($^3P$) + Ar atoms. 
For each electronic state,  $V(R)$ is obtained from 
{\it ab initio} calculations. The potential  of the $(1)^3\Sigma^+$
state is taken from Ref.~\cite{HapkaJCP13}. For the states resulting
from the interaction of argon with  He($^3P$),  we have calculated
the relevant  potential energy curves by using the spin-restricted  
open-shell coupled cluster method  restricted to single,   double,  and
noniterative triple excitations [RCCSD(T)]. In the initial restricted
open-shell Hartree-Fock (ROHF) calculations, we have alternated the
occupied orbitals in order to force convergence of the ROHF wave
function to the desired excited electronic state. A similar approach
was successfully used to describe the interaction between  
He($^3S,{^3P}$) and H$_2$~\cite{HapkaJCP13, ShagamNatChem15}.  Here,
we have  employed a doubly augmented d-aug-cc-pVQZ  basis set 
for the helium atom and a singly augmented aug-cc-pVQZ basis set for
the argon atom~\cite{Dunning}.
The electronic structure calculations were performed with the MOLPRO
suite of codes~\cite{molpro10}.
The long-range part of $V(R)$, for $R>20\,$bohr was taken in its
asymptotic form, $C_6/R^6$ accounting only for the leading order term,
and fixing  the van der Waals $C_6$ coefficients to 214.1$\,$a.u.  for
the $(1)^3\Sigma^+$  state~\cite{HapkaJCP13}, 369.2 a.u. for the
$(2)^3\Sigma^+$ state, and 201.9 for the $^3\Pi$ state.  
To calculate the last two $C_6$ coefficients, the  polarizability for
He($^3P$) was constructed from the sum-over-states expression as
described in Ref.~\cite{ShagamNatChem15}. 

For the imaginary part,  we have employed the functional form of
$\Gamma(R)$ from Ref.~\cite{width}, 
\begin{equation}
\Gamma(R) = A e^{-\alpha R} + B e^{-\beta R} \,,
  \label{eq:Gamma}
\end{equation}
where the original values of the parameters, $A=896\,$eV,
$B=0.0043\,$meV, $\alpha = 4.08\,$\AA$^{-1}$ and $\beta =
0.28\,$\AA$^{-1}$, were obtained by a fitting procedure to 
reproduce the Penning ionization cross sections for collisions
between He($^3S$) and Ar at thermal energies~\cite{width}. 
We have assumed $\Gamma(R)$ to be identical for all relevant
electronic states. This assumption can be justified on the grounds  of
recent results for He($^3S,^3P$) + H$_2$~\cite{ShagamNatChem15}, where
calculations have shown  $\Gamma(R)$  to be almost the same  for  
He($^3S$) + H$_2$  and  He($^3P$) + H$_2$. 

The Hamiltonian  for the nuclear motion in 
each electronic state $n$ takes the following form:
\begin{equation}
  \widehat{H}_{n} =  \frac{1}{2\mu}
  \frac{\partial^2 } {\partial R^2} 
  +\frac{l(l+1)}{2\mu R^2} +V_{n}(R) + \frac{i}{2} \; \Gamma(R)
\end{equation}
with $\mu$ denoting the reduced mass, and $l$  the rotational quantum
number. In the rotational kinetic energy, we have neglected the
spin-rotational coupling  
and nonadiabatic coupling between $^3\Sigma$ and $^3\Pi$ states. These
simplifications are justified since we are mostly interested in collision
energies of the order of 1 Kelvin and relatively high  partial waves,
$l \sim 7$, 
where the rotational Hamiltonian is dominated by the $l(l+1)$ term, so
$l$ is a good quantum number. 
Diagonalization of this  Hamiltonian for each $n$ and $l$ gives  the
bound vibrational levels and continuum states. The presence of the
optical potential implies
complex eigenvalues, $W_{v,l} =  E_{v,l} + i/2 \;\Gamma_{v,l}^{p}$  
where $E_{v,l}$ denotes the position of the resonance and
$\Gamma_{v,l}^{p}$ is  the corresponding width due to Penning ionization
process. 

Photoassociation rate coefficients $K_{PA}$  for  collision energies
$E$  and a laser frequency  $\omega$ are obtained from the standard
expression~\cite{kpa1, kpa2},  
\begin{equation}
  K_{PA}(E, \omega)=\frac{\pi \hbar}{\mu k} \sum_l (2l+1) \sum_{v', l'}
  |S_{v',l'}(E, \omega, l)|^2\,,
\end{equation}
where $k=\sqrt{2\mu E/ \hbar^2}$, adapted to Penning ionization. The
transition probability $|S_{v',l'}(E, \omega, l)|^2$ for
photoassociation from a continuum state with energy $E$ 
just above the He$(^3S)$+Ar asymptote into a bound
level with quantum numbers $v',l'$  just below the dissociation limit
He$(^3P)$+Ar reads
\begin{equation}
  |S_{v',l'}(E, \omega, l)|^2 = \frac{\Gamma^s_{v',l'}(E,l) \cdot \Gamma^d_{v',l'}}{(E-\Delta_{v',l'})^2 +1/4 [\Gamma^s_{v',l'}(E,l) + \Gamma^d_{v',l'}]^2}\,,
\end{equation}
where $ \Gamma^s_{v',l'}(E,l) $ is the stimulated emission rate, $ \Gamma^d_{v',l'}$
the total decay rate, and  $\Delta_{v',l'} = E_{v',l'} - \hbar
\omega$ the detuning of the laser frequency from the vibrational
level. The total decay rate  $ \Gamma^d_{v',l'}$ is the sum 
of the natural radiative width $\Gamma^n_{v',l'}$ and the 
Penning ionization width $\Gamma_{v',l'}^p$. At low laser intensity
$I$, the stimulated emission rate  $\Gamma^s_{v',l'}(E,l)$ is given by
Fermi's golden rule,  
\begin{equation}
 \Gamma^s_{v',l'}(E,l) = \frac{4 \pi^2 I}{4 \pi \epsilon_0 c} (2l'+1) H_{l'} |\langle \psi_{E,l}(R)|d(R)| \psi_{v',l'} \rangle|^2\,,
\end{equation}
where $H_{l'}$ is the H\"{o}ln-London factor equal to $(l'+1)/(2l'+1)$
for $l'=l-1$, to $1/(2l'+1)$ for $l=l'$,  and to $l'/(2l'+1)$ for
$l'=l+1$.  
$\psi_{E,l}(R)$ denotes the continuum wave function  for energy $E$
and partial  wave $l$, obtained from diagonalization of the
Hamiltonian for the  $(1)^3\Sigma^+$ 
state with a very large grid, while $ \psi_{v',l'}(R)$  is the bound state
wave function of either the $(1)^3\Sigma^+$ or  $^3\Pi$
electronic state. The molecular transition  dipole moments $d(R)$ from
$(1)^3\Sigma^+$ to  $(2)^3\Sigma^+$ and from   $(1)^3\Sigma^+$ to
$^3\Pi$ electronic states were assumed to be constant and equal to the
atomic transition dipole moment $^3S \leftarrow {^3}P$ in He, which is
2.53$\,$a.u. 
Similarly, a constant transition dipole moment $d$ was assumed in the 
calculations of the natural radiative widths $\Gamma^n_{v',l'}$. 
We have not performed any thermal averaging to calculate $K_{PA}(E)$
since the experimental conditions with merged beams allow for probing
a particular collision energy with essentially no thermal 
broadening. In all photoassociation calculations we have assumed a
laser intensity of 1$\;$W/cm$^2$.  

\section{Results}
\label{sec:results}
\begin{figure}[tb]
   \includegraphics[width=0.5\linewidth, angle=-90]{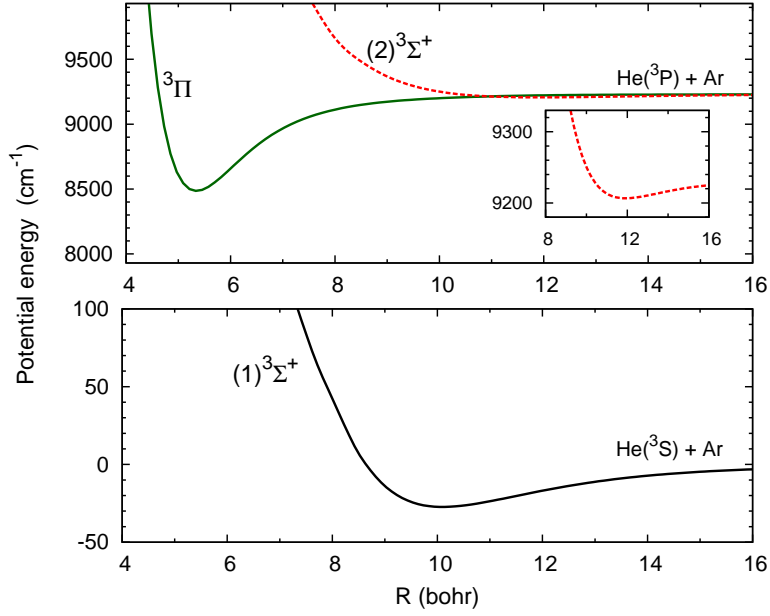}
   \caption{Potential energy curves  for the $(1)^3\Sigma^+$,
     $(2)^3\Sigma^+$ and  $^3\Pi$ electronic states. 
     Zero interaction energy  is fixed at the He$(^3S)$ + Ar atomic asymptote.
  }
  \label{fig1}
\end{figure}
Figure~\ref{fig1} presents the three potential energy curves $V(R)$ of
the helium-argon collision complex relevant for photoassociation in
Penning ionization.  
Both the $(1)^3\Sigma^+$ and $(2)^3\Sigma^+$ states are rather shallow 
with their  minima found at relatively large interatomic
distances. More precisely, the $(1)^3\Sigma^+$ state has a well depth
of 27.3$\,$cm$^{-1}$ with its equilibrium  distance  equal  to
10.09$\,$bohr, while the $(2)^3\Sigma^+$ state has its minimum of
24.3$\,$cm$^{-1}$ at 11.92$\,$bohr.  A qualitative difference  is
found for the potential energy curve of the  $^3\Pi$ state, which has
a predicted minimum of 
745$\,$cm$^{-1}$  at 5.34$\,$bohr. This pattern of the electronic  states,
with weakly bound  $\Sigma$ states and a strongly interacting  $\Pi$ state
is fully analogous  with  the previously reported interaction
potentials for He($^3S,^3P$) colliding with H$_2$
molecules~\cite{ShagamNatChem15} or other similar systems  where a
single-valence atom interacts  with closed-shell species, such as
Li($^2S,^2P$) + H$_2$~\cite{potLiH2}. 

The  entrance channel  for photoassociation spectroscopy of the
Penning ionization reaction is the $(1)^3\Sigma^+$  state, resulting
from the interaction between metastable He($^3S$) 
and ground-state Ar atoms. We consider a detuning of the
photoassociation laser with respect to the  $^3P \; \leftarrow {^3S}$
helium atomic line, so the targeted  bound levels belong to either the 
$(2)^3\Sigma^+$  or the $^3\Pi$ electronic states. In our model,  we
neglect any nonadiabatic  (spin-rotation or spin-orbit) couplings
between the $(2)^3\Sigma^+$  and  $^3\Pi$ states. These couplings should
be important only for ultralow collisional energies and for the most 
weakly bound levels, where the $(2)^3\Sigma^+$  and $^3\Pi$ states
are nearly degenerate. For more deeply bound levels and for sub-Kelvin
or higher collision energies,  
the contributions from any nonadiabatic coupling would be of the order or
smaller than the accuracy of the electronic potentials. Thus we can
safely neglect them without compromising the essential features  of
our model. 

\begin{figure}[tb]
   \includegraphics[width=0.5\linewidth, angle=-90]{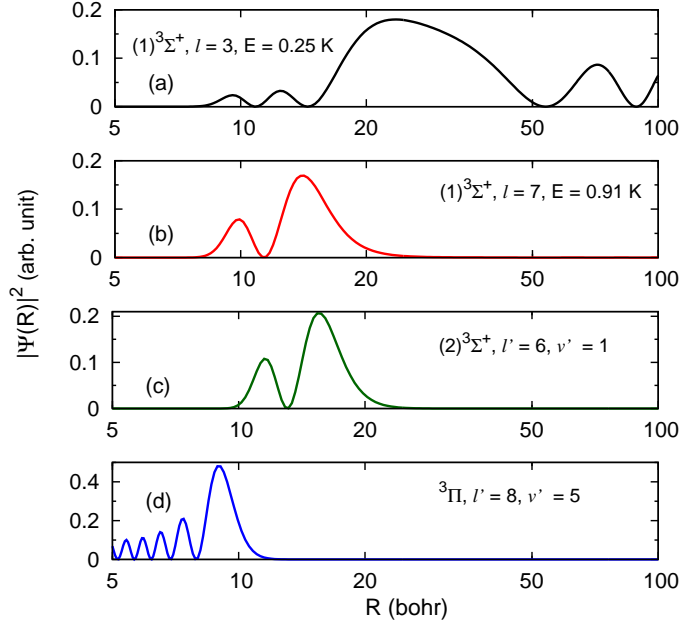}
   \caption{(a), (b) Wave functions for two considered shape resonances in
     $(1)^3\Sigma^+$ state. (c), (d) Wave functions for bound levels
     in  $(2)^3\Sigma^+$ and  $^3\Pi$ electronic states particularly suited
     for photoassociation starting from the $l=7$ shape resonance.
  }
  \label{fig2}
\end{figure}
Low-energy collisions between He($^3S$) and Ar are characterized by
several shape resonances, as recently demonstrated in an experiment
with merged beams~\cite{HensonSci12}. The lowest two  resonances are
located at approximately  0.23$\,$Kelvin and 1.10$\,$Kelvin.
The positions of these shape resonances are comparatively well
reproduced by our model---they occur  at collisional energies of
0.25$\,$Kelvin (partial wave $l=3$) and 0.91$\,$Kelvin   
(partial wave $l=7$). 
Photoassociation at energies corresponding to a shape resonance is 
highly efficient~\cite{GonzalezPRA12}:  The enhanced probability  for
transitions from a scattering state to a bound level is due to a
largely increased amplitude of resonance wave functions. In other
words, for an orbiting resonance, amplitude becomes trapped at short 
inter-particle separations, inside the rotational barrier, leading to
larger transition matrix elements in photoassociation. To take
advantage of this enhancement, we focus on photoassociation at
collisional energies matching the theoretical 
positions of the lowest shape resonances, i.e., 0.25$\,$Kelvin  and
0.91$\,$Kelvin. Somewhat inexpectedly, 
the nature of these two shape resonances is quite different. This is
illustrated by the two upper panels of Fig.~\ref{fig2}, showing the
wave functions of these resonances. The energy of the $l=3$ resonance
is slightly above the maximum 
of the centrifugal barrier which is equal to 0.23$\,$Kelvin. As a
result, the wave function is fairly diffuse with only little enhancement
of pair density  at short inter-particle 
distances. In contrast, the nature of  the $l=7$ shape resonance
[Fig.\ref{fig2}(b)] is very different. It is energetically located
deeply inside the well created by the centrifugal 
barrier. Thus the amplitude of its wave function is enhanced at short
inter-particle separations, resembling at these distances the  wave
function of a typical bound level.

\begin{figure}[tb]
   \includegraphics[width=0.5\linewidth, angle=-90]{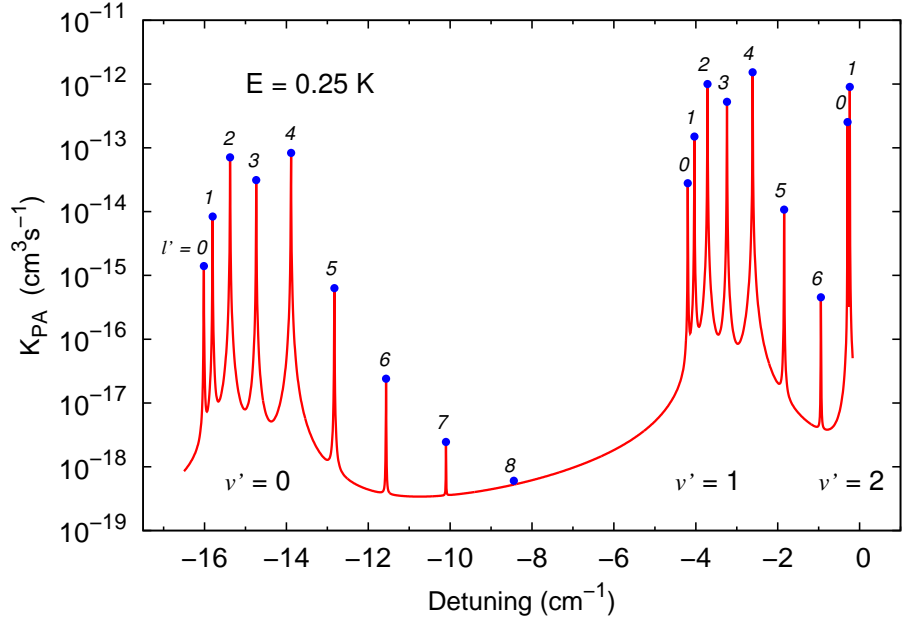}
  \caption{Photoassociation rates as a function of laser detuning from
    the He $^3S\to^3P$ atomic line for rovibrational levels of the 
    $(2)^3\Sigma^+$ state at a collision energy of 0.25$\,$Kelvin and
    a laser intensity of $1\;$W/cm$^2$. The vibrational and rotational
    quantum numbers are indicated for each peak. The peaks for
    $l'=2,3,4$ are enhanced due to the $l=3$ shape resonance as 
    initial state at 0.25$\,$K, whereas for all other peaks
    photoassociation starts from regular scattering states.
  }
  \label{fig3}
\end{figure}
Photoassociation rates for levels in the $(1)^3\Sigma^+$  and
$^3\Pi$ electronic states are obtained separately   
from single-channel calculations since we neglect nonadiabatic
couplings. Figure~\ref{fig3}
presents the spectrum for photoassociation into
the $(2)^3\Sigma^+$ state at a collision energy of 0.25$\,$Kelvin.
A rotational progression for three vibrational levels is clearly
visible. Note that our potential energy curves are not sufficiently
accurate to predict the exact detunings of these peaks. Refinement of
the potentials based on spectroscopic information would be required to
this end. However, the spacing between rotational and vibrational
levels can be predicted correctly. 

The highest rates in Fig.~\ref{fig3} occur for $2\le l'\le 4$, i.e., for
bound levels which are directly accessible 
from the  $l=3$ shape resonance. However, the observed enhancement of
the rates compared to the other levels 
is  one order of magnitude at most. This  relatively minor  resonance 
effect is easily explained by the above-the-barrier character of this
shape resonance, and the corresponding diffuse nature of its
wavefunction, cf. Fig.~\ref{fig2}(a).
The widths of the photoassociation lines
range from 30$\,$MHz (for $v'=2$) up to 160$\,$MHz (for $v'=0$), 
and are almost purely determined by the Penning ionization
contributions $\Gamma^p_{v',l'}$, since the width due to spontaneous
radiative emission $\Gamma^n_{v',l'}$  is only  of the order of 1$\,$MHz.  
Increase of the peak width with binding energy is readily explained by
the stronger localization of the bound level amplitude at shorter
inter-particle separation together with the exponential increase of
the ionization probability with decreasing inter-particle separation,
cf. Eq.~\eqref{eq:Gamma}. More generally, this also explains why the
bound levels ionize faster than scattering states, leading to an
observable signature of photoassociation in the Penning ionization
reaction. 

\begin{figure}[tb]
   \includegraphics[width=0.5\linewidth, angle=-90]{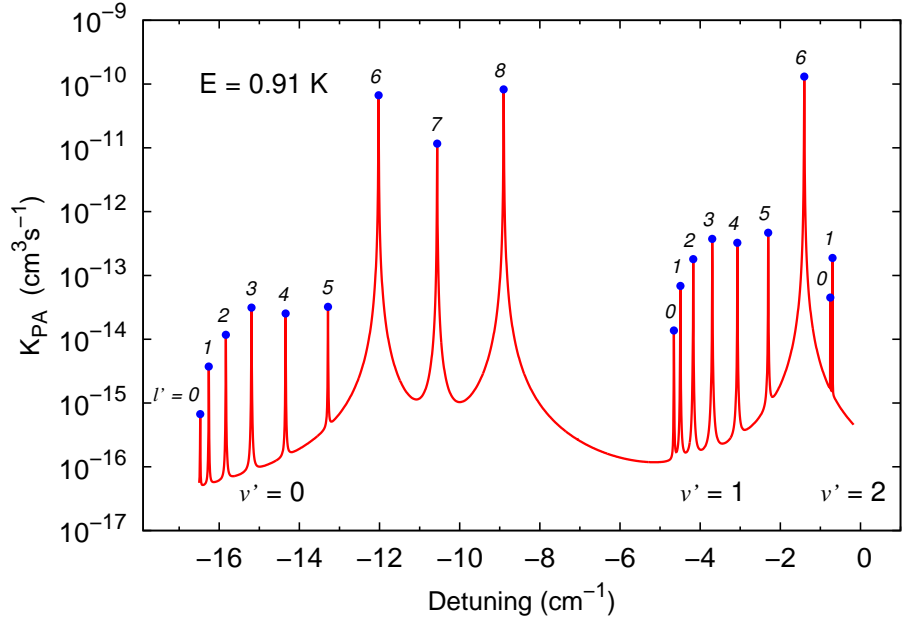}
  \caption{Photoassociation rates as a function of laser detuning from
    the He $^3S\to^3P$ atomic line for rovibrational levels of the 
    $(2)^3\Sigma^+$ state at a collision energy of 0.91$\,$Kelvin and
    a laser intensity of $1\;$W/cm$^2$. The vibrational and rotational
    quantum numbers are indicated for each peak.  The peaks for
    $l'=6,7,8$ are enhanced due to the $l=7$ shape resonance as 
    initial state at 0.91$\,$K, whereas for all other peaks
    photoassociation starts from regular scattering states.
  }
  \label{fig4}
\end{figure}
The photoassociation spectrum for bound levels in the $(2)^3\Sigma^+$
state at a collision  energy  
of 0.91$\,$Kelvin is presented in Fig.~\ref{fig4}. Here, strongly
enhanced  peaks are observed at detunings matching the  rovibational
levels  with $6\le l'\le 8$. These levels can be directly accessed
from the $l=7$ shape resonance at 0.91$\,$Kelvin. The enhancement due
to the shape resonance is nearly three orders of magnitude, 
reaching a peak rate of 10$^{-10}\;$cm$^3$/s for the assumed laser
intensity of $I=1\;$W/cm$^2$, much more than in the previous case of
the $l=3$ resonance. This strong enhancement is easily rationalized by
comparing the wave functions for  the $l=7$ 
shape resonance and the $l'=6$, $(2)^3\Sigma^+$ bound level, shown in
panels (b) and (c) of Fig.~\ref{fig2}: The  
very good  overlap between these wave functions leads to the predicted
large photoassociation rate. 

Overall, Figs.~\ref{fig3} and~\ref{fig4} predict, for levels in the
$(2)^3\Sigma^+$ electronic state, photoassociation
spectra  with a well resolved rotational-vibrational
structure. Penning ionization broadens the peaks compared to the
natural linewidth. However, this broadening  is significantly  smaller
than the spacing between the lines.  
Thus, photoassociation spectroscopy is expected to be feasible with
high resolution and can
be used, for example, to determine the rotational quantum number of
the shape resonance that serves as starting point. The basic feature
of the potential energy curve which allows for this prediction of
well-resolved photoassociation lines is
a potential well at comparatively large inter-particle separation where
the probability for Penning ionization is not very large.

\begin{figure}[tb]
   \includegraphics[width=0.5\linewidth, angle=-90]{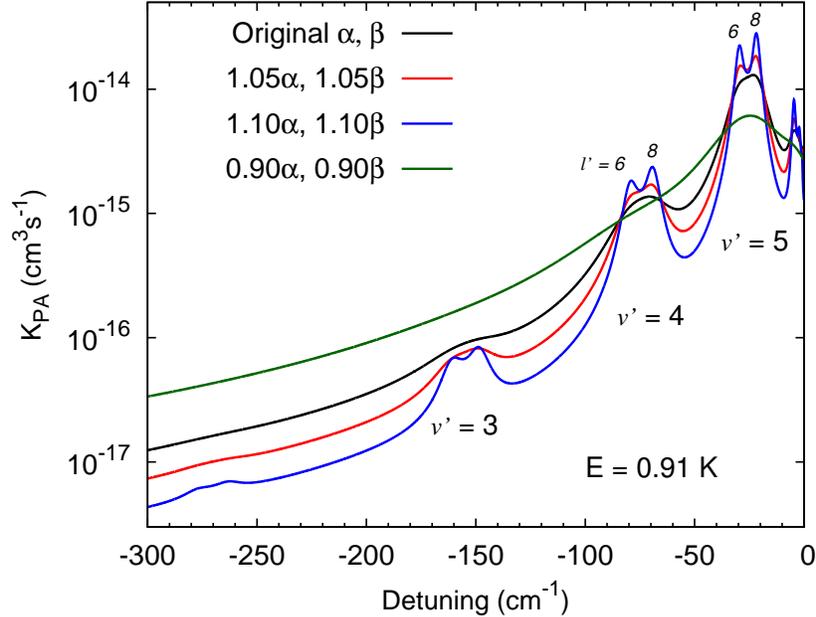}
  \caption{Photoassociation rates as a function of laser detuning from
    the He $^3S\to^3P$ atomic line for rovibrational levels of the 
    $^3\Pi$ state at a collision energy of 0.91$\,$Kelvin and
    a laser intensity of $1\;$W/cm$^2$. The vibrational and rotational
    quantum numbers are indicated for each peak. $\alpha$ and $\beta$
    are the parameters of the optical potential, Eq.~\eqref{eq:Gamma},
    with the original values taken from Ref.~\cite{width}. 
  }
  \label{fig5}
\end{figure}
Finally, we consider photoassociation  into  the $^3\Pi$ state. We
focus only on a collision energy 
of 0.91$\,$Kelvin, where the sharpest resonance effect and thus the
clearest signal is expected. The 
photoassociation spectrum, obtained with the original parameters of
the imaginary potential, $\alpha$ and $\beta$ in Eq.~\eqref{eq:Gamma},
taken from Ref.~\cite{width},  is presented by the  
black curve in Fig.~\ref{fig5}. In contrast to the $(2)^3\Sigma^+$
state,  the rovibrational structure of the spectrum  has almost
disappeared. For bound levels in the $^3\Pi$ state, the Penning  
ionization widths are  in the range from 1$\,$cm$^{-1}$ or 30$\,$GHz
(for the weakly bound $v'=7$ level) up to 60$\,$cm$^{-1}$  
(for the most deeply bound $v'=0$ level). These large widths,
comparable to the spacings between the rovibrational levels, do not
allow to resolve individual lines in the spectrum, in particular for
larger detunings, i.e., larger binding energies. This raises the
question how the strength of the imaginary potential, or more
precisely whether and how the parameters $\alpha$ and
$\beta$ in Eq.~\eqref{eq:Gamma}, influence the
photoassociation spectrum. To answer this question, 
we scale the exponents in Eq.~\eqref{eq:Gamma} by $\pm10$\%.
The resulting photoassociation spectra are also depicted in
Fig.~\ref{fig5}.  If $\Gamma(R)$ is being weakened (larger
$\alpha$ and $\beta$, red and blue lines in Fig.~\ref{fig5}), 
individual rovibrational lines emerge.  On the other hand, the 
spectrum becomes almost flat for stronger $\Gamma(R)$ (green line in
Fig.~\ref{fig5}), as expected: If the ionization probability is very
high, a wave packet cannot oscillate back and forth in the potential
even once, i.e., the discrete nature of the bound levels cannot be
established. Thus, for a potential energy curve located at
shorter inter-particle separations, the prospects for
photoassociation spectroscopy depend highly on the actual strength of
$\Gamma(R)$. In turn,  photoassociation spectroscopy can be used to 
determine this strength. If the ionization probability is not too
high, such that single rovibrational peaks can be resolved, the
increase of peak widths with binding energy should moreover allow to
determine the functional form of $\Gamma(R)$. In other words, 
photoassociation spectra as those represented by the blue line 
in Fig.~\ref{fig5} can be used to fully map out $\Gamma(R)$, a
quantity that still poses a rather large 
challenge to {\it ab initio} calculations.

\section{Conclusions}
\label{sec:concl}

We have studied the prospects for photoassociation as a spectroscopic
tool to study Penning ionization reactions. Introducing an \textit{ab
  initio} model for the simplest example, Penning ionization of
metastable helium with argon, we find usability of this tool to be
determined by the position of the excited state well. For a potential
with a comparatively large equilibrium distance, about 10$\,$bohr in our
case, the probability of Penning ionization is larger than in the
entrance channel, yet sufficiently small to allow for resolving single
rovibrational levels. Photoassociation can then be observed by peaks
in the ionization rate as a function of laser detuning. 

In merged
beams, the collision energy in the entrance channel can precisely be
controlled. Choosing in this way to photoassociate out of a shape
resonance allows for unequivocally determining the rotational quantum
number of the resonance since, compared to regular scattering states, a
shape resonance leads to clearly enhanced photoassociation peaks in 
the rotational progression. While the enhancement amounts to about one
order of magnitude for diffuse, above-the-barrier shape resonances,
enhancements up to three orders of magnitude are expected for shape
resonances whose energy is well below the top of the rotational
barrier.

For a potential with a comparatively small equilibrium distance, about
5$\,$bohr in our case, the probability of Penning ionization becomes
very large. Whether photoassociation peaks as a function of laser
detuning can be observed or not, then depends very sensitively on the
value of the exponents in the imaginary potential. We have started our
calculations with values from the literature, obtained by fitting
experimental data obtained for thermal collision
energies~\cite{width}. Changing these values by only -10 per cent
completely washes out any remnants of a peak structure, whereas a
change in +10 per cent leads to clearly observable photoassociation
peaks. Photoassociation spectroscopy can thus be used to accurately
gauge the exponents of the imaginary potential. Moreover, if the
Penning ionization probability is sufficiently small to allow for peak
widths to be determined, an increase of the peak width with binding
energy, respectively laser detuning, can be used to fully map out the
functional form of the imaginary potential. 

In summary, we find photoassociation spectroscopy to be an extremely
useful tool to obtain in-depth information on Penning ionization
reactions, provided the ionization proceeds not too fast. Our
conclusions should hold also for collision partners other than argon,
in particular if these are closed-shell.

\begin{acknowledgments}
  Financial support from the German-Israeli Foundation, grant
  no. 1254, is gratefully acknowledged. 
\end{acknowledgments}


\end{document}